\documentclass[reprint,aps,prd,twocolumn,superscriptaddress,showpacs,preprintnumbers,
               amsmath,amssymb,floatfix]{revtex4-1}%

\usepackage{graphicx}% Include figure files
\usepackage{epstopdf}

\usepackage[mathlines]{lineno}% Enable numbering of text and display math
\usepackage{xspace} % for abbreviations

\newcommand{\eg}{{\it e.g.}\xspace}

\def\gev{GeV/$c^2$\xspace}

\def\kms{km/s\xspace}
\def\ba{$^{133}$Ba\xspace}
\def\cf{$^{252}$Cf\xspace}
\def\pb{$^{206}$Pb\xspace}

\begin{document}
\title{Silicon Detector Dark Matter Results from the Final Exposure of CDMS II}

%
%%Put affiliations first to force them into alphabetical order
%
\affiliation{Division of Physics, Mathematics, \& Astronomy, California Institute of Technology, Pasadena, CA 91125, USA}
\affiliation{Fermi National Accelerator Laboratory, Batavia, IL 60510, USA}
\affiliation{Lawrence Berkeley National Laboratory, Berkeley, CA 94720, USA}
\affiliation{Department of Physics, Massachusetts Institute of Technology, Cambridge, MA 02139, USA}
\affiliation{Pacific Northwest National Laboratory, Richland, WA 99352, USA}
\affiliation{Department of Physics, Queen's University, Kingston ON, Canada K7L 3N6}
\affiliation{Department of Physics, Santa Clara University, Santa Clara, CA 95053, USA}
\affiliation{SLAC National Accelerator Laboratory/Kavli Institute for Particle Astrophysics and Cosmology, 2575 Sand Hill Road, Menlo Park 94025, CA}
\affiliation{Department of Physics, Southern Methodist University, Dallas, TX 75275, USA}
\affiliation{Department of Physics, Stanford University, Stanford, CA 94305, USA}
\affiliation{Department of Physics, Syracuse University, Syracuse, NY 13244, USA}
\affiliation{Department of Physics, Texas A\&M University, College Station, TX 77843, USA}
\affiliation{Departamento de F\'{\i}sica Te\'orica and Instituto de F\'{\i}sica Te\'orica UAM/CSIC, Universidad Aut\'onoma de Madrid, 28049 Madrid, Spain}
\affiliation{Department of Physics, University of California, Berkeley, CA 94720, USA}
\affiliation{Department of Physics, University of California, Santa Barbara, CA 93106, USA}
\affiliation{Department of Physics, University of Colorado, Denver, CO 80217, USA}
\affiliation{Department of Physics, University of Evansville, Evansville, IN 47722, USA}
\affiliation{Department of Physics, University of Florida, Gainesville, FL 32611, USA}
\affiliation{School of Physics \& Astronomy, University of Minnesota, Minneapolis, MN 55455, USA}
\affiliation{Physics Institute, University of Z\"{u}rich, Winterthurerstr. 190, CH-8057, Switzerland}
\author{R.~Agnese} \affiliation{Department of Physics, University of Florida, Gainesville, FL 32611, USA}
\author{Z.~Ahmed} \affiliation{Division of Physics, Mathematics, \& Astronomy, California Institute of Technology, Pasadena, CA 91125, USA}
\author{A.J.~Anderson} \affiliation{Department of Physics, Massachusetts Institute of Technology, Cambridge, MA 02139, USA}
\author{S.~Arrenberg} \affiliation{Physics Institute, University of Z\"{u}rich, Winterthurerstr. 190, CH-8057, Switzerland}
\author{D.~Balakishiyeva} \affiliation{Department of Physics, University of Florida, Gainesville, FL 32611, USA}
\author{R.~Basu~Thakur~} \affiliation{Fermi National Accelerator Laboratory, Batavia, IL 60510, USA}
\author{D.A.~Bauer} \affiliation{Fermi National Accelerator Laboratory, Batavia, IL 60510, USA}
\author{J.~Billard} \affiliation{Department of Physics, Massachusetts Institute of Technology, Cambridge, MA 02139, USA}
\author{A.~Borgland} \affiliation{SLAC National Accelerator Laboratory/Kavli Institute for Particle Astrophysics and Cosmology, 2575 Sand Hill Road, Menlo Park 94025, CA}
\author{D.~Brandt} \affiliation{SLAC National Accelerator Laboratory/Kavli Institute for Particle Astrophysics and Cosmology, 2575 Sand Hill Road, Menlo Park 94025, CA}
\author{P.L.~Brink} \affiliation{SLAC National Accelerator Laboratory/Kavli Institute for Particle Astrophysics and Cosmology, 2575 Sand Hill Road, Menlo Park 94025, CA}
\author{T.~Bruch} \affiliation{Physics Institute, University of Z\"{u}rich, Winterthurerstr. 190, CH-8057, Switzerland}
\author{R.~Bunker} \affiliation{Department of Physics, Syracuse University, Syracuse, NY 13244, USA}
\author{B.~Cabrera} \affiliation{Department of Physics, Stanford University, Stanford, CA 94305, USA}
\author{D.O.~Caldwell} \affiliation{Department of Physics, University of California, Santa Barbara, CA 93106, USA}
\author{D.G.~Cerdeno} \affiliation{Departamento de F\'{\i}sica Te\'orica and Instituto de F\'{\i}sica Te\'orica UAM/CSIC, Universidad Aut\'onoma de Madrid, 28049 Madrid, Spain}
\author{H.~Chagani} \affiliation{School of Physics \& Astronomy, University of Minnesota, Minneapolis, MN 55455, USA}
\author{J.~Cooley} \affiliation{Department of Physics, Southern Methodist University, Dallas, TX 75275, USA}
\author{B.~Cornell} \affiliation{Division of Physics, Mathematics, \& Astronomy, California Institute of Technology, Pasadena, CA 91125, USA}
\author{C.H.~Crewdson} \affiliation{Department of Physics, Queen's University, Kingston ON, Canada K7L 3N6}
\author{P.~Cushman} \affiliation{School of Physics \& Astronomy, University of Minnesota, Minneapolis, MN 55455, USA}
\author{M.~Daal} \affiliation{Department of Physics, University of California, Berkeley, CA 94720, USA}
\author{F.~Dejongh} \affiliation{Fermi National Accelerator Laboratory, Batavia, IL 60510, USA}
%\author{P.C.F.~Di~Stefano} \affiliation{Department of Physics, Queen's University, Kingston ON, Canada K7L 3N6}
\author{E.~Do~Couto~E~Silva} \affiliation{SLAC National Accelerator Laboratory/Kavli Institute for Particle Astrophysics and Cosmology, 2575 Sand Hill Road, Menlo Park 94025, CA}
\author{T.~Doughty} \affiliation{Department of Physics, University of California, Berkeley, CA 94720, USA}
\author{L.~Esteban} \affiliation{Departamento de F\'{\i}sica Te\'orica and Instituto de F\'{\i}sica Te\'orica UAM/CSIC, Universidad Aut\'onoma de Madrid, 28049 Madrid, Spain}
\author{S.~Fallows} \affiliation{School of Physics \& Astronomy, University of Minnesota, Minneapolis, MN 55455, USA}
\author{E.~Figueroa-Feliciano} \affiliation{Department of Physics, Massachusetts Institute of Technology, Cambridge, MA 02139, USA}
\author{J.~Filippini} \affiliation{Division of Physics, Mathematics, \& Astronomy, California Institute of Technology, Pasadena, CA 91125, USA}
\author{J.~Fox} \affiliation{Department of Physics, Queen's University, Kingston ON, Canada K7L 3N6}
\author{M.~Fritts} \affiliation{School of Physics \& Astronomy, University of Minnesota, Minneapolis, MN 55455, USA}
\author{G.L.~Godfrey} \affiliation{SLAC National Accelerator Laboratory/Kavli Institute for Particle Astrophysics and Cosmology, 2575 Sand Hill Road, Menlo Park 94025, CA}
\author{S.R.~Golwala} \affiliation{Division of Physics, Mathematics, \& Astronomy, California Institute of Technology, Pasadena, CA 91125, USA}
\author{J.~Hall} \affiliation{Pacific Northwest National Laboratory, Richland, WA 99352, USA}
\author{R.H.~Harris} \affiliation{Department of Physics, Texas A\&M University, College Station, TX 77843, USA}
\author{S.A.~Hertel} \affiliation{Department of Physics, Massachusetts Institute of Technology, Cambridge, MA 02139, USA}
\author{T.~Hofer} \affiliation{School of Physics \& Astronomy, University of Minnesota, Minneapolis, MN 55455, USA}
\author{D.~Holmgren} \affiliation{Fermi National Accelerator Laboratory, Batavia, IL 60510, USA}
\author{L.~Hsu} \affiliation{Fermi National Accelerator Laboratory, Batavia, IL 60510, USA}
\author{M.E.~Huber} \affiliation{Department of Physics, University of Colorado, Denver, CO 80217, USA}
\author{A.~Jastram} \affiliation{Department of Physics, Texas A\&M University, College Station, TX 77843, USA}
\author{O.~Kamaev} \affiliation{Department of Physics, Queen's University, Kingston ON, Canada K7L 3N6}
\author{B.~Kara} \affiliation{Department of Physics, Southern Methodist University, Dallas, TX 75275, USA}
\author{M.H.~Kelsey} \affiliation{SLAC National Accelerator Laboratory/Kavli Institute for Particle Astrophysics and Cosmology, 2575 Sand Hill Road, Menlo Park 94025, CA}
\author{A.~Kennedy} \affiliation{School of Physics \& Astronomy, University of Minnesota, Minneapolis, MN 55455, USA}
\author{P.~Kim} \affiliation{SLAC National Accelerator Laboratory/Kavli Institute for Particle Astrophysics and Cosmology, 2575 Sand Hill Road, Menlo Park 94025, CA}
\author{M.~Kiveni} \affiliation{Department of Physics, Syracuse University, Syracuse, NY 13244, USA}
\author{K.~Koch} \affiliation{School of Physics \& Astronomy, University of Minnesota, Minneapolis, MN 55455, USA}
\author{M.~Kos} \affiliation{Department of Physics, Syracuse University, Syracuse, NY 13244, USA}
\author{S.W.~Leman} \affiliation{Department of Physics, Massachusetts Institute of Technology, Cambridge, MA 02139, USA}
\author{B.~Loer} \affiliation{Fermi National Accelerator Laboratory, Batavia, IL 60510, USA}
\author{E. Lopez~Asamar}  \affiliation{Departamento de F\'{\i}sica Te\'orica and Instituto de F\'{\i}sica Te\'orica UAM/CSIC, Universidad Aut\'onoma de Madrid, 28049 Madrid, Spain}
\author{R.~Mahapatra} \affiliation{Department of Physics, Texas A\&M University, College Station, TX 77843, USA}
\author{V.~Mandic} \affiliation{School of Physics \& Astronomy, University of Minnesota, Minneapolis, MN 55455, USA}
\author{C.~Martinez} \affiliation{Department of Physics, Queen's University, Kingston ON, Canada K7L 3N6}
\author{K.A.~McCarthy} \affiliation{Department of Physics, Massachusetts Institute of Technology, Cambridge, MA 02139, USA}
\author{N.~Mirabolfathi} \affiliation{Department of Physics, University of California, Berkeley, CA 94720, USA}
\author{R.A.~Moffatt} \affiliation{Department of Physics, Stanford University, Stanford, CA 94305, USA}
\author{D.C.~Moore} \affiliation{Division of Physics, Mathematics, \& Astronomy, California Institute of Technology, Pasadena, CA 91125, USA}
\author{P.~Nadeau} \affiliation{Department of Physics, Queen's University, Kingston ON, Canada K7L 3N6}
\author{R.H.~Nelson} \affiliation{Division of Physics, Mathematics, \& Astronomy, California Institute of Technology, Pasadena, CA 91125, USA}
\author{K.~Page} \affiliation{Department of Physics, Queen's University, Kingston ON, Canada K7L 3N6}
\author{R.~Partridge} \affiliation{SLAC National Accelerator Laboratory/Kavli Institute for Particle Astrophysics and Cosmology, 2575 Sand Hill Road, Menlo Park 94025, CA}
\author{M.~Pepin} \affiliation{School of Physics \& Astronomy, University of Minnesota, Minneapolis, MN 55455, USA}
\author{A.~Phipps} \affiliation{Department of Physics, University of California, Berkeley, CA 94720, USA}
\author{K.~Prasad} \affiliation{Department of Physics, Texas A\&M University, College Station, TX 77843, USA}
\author{M.~Pyle} \affiliation{Department of Physics, University of California, Berkeley, CA 94720, USA}
\author{H.~Qiu} \affiliation{Department of Physics, Southern Methodist University, Dallas, TX 75275, USA}
\author{W.~Rau} \affiliation{Department of Physics, Queen's University, Kingston ON, Canada K7L 3N6}
\author{P.~Redl} \affiliation{Department of Physics, Stanford University, Stanford, CA 94305, USA}
\author{A.~Reisetter} \affiliation{Department of Physics, University of Evansville, Evansville, IN 47722, USA}
\author{Y.~Ricci} \affiliation{Department of Physics, Queen's University, Kingston ON, Canada K7L 3N6}
\author{T.~Saab} \affiliation{Department of Physics, University of Florida, Gainesville, FL 32611, USA}
\author{B.~Sadoulet} \affiliation{Department of Physics, University of California, Berkeley, CA 94720, USA}\affiliation{Lawrence Berkeley National Laboratory, Berkeley, CA 94720, USA}
\author{J.~Sander} \affiliation{Department of Physics, Texas A\&M University, College Station, TX 77843, USA}
\author{K.~Schneck} \affiliation{SLAC National Accelerator Laboratory/Kavli Institute for Particle Astrophysics and Cosmology, 2575 Sand Hill Road, Menlo Park 94025, CA}
\author{R.W.~Schnee} \affiliation{Department of Physics, Syracuse University, Syracuse, NY 13244, USA}
\author{S.~Scorza} \affiliation{Department of Physics, Southern Methodist University, Dallas, TX 75275, USA}
\author{B.~Serfass} \affiliation{Department of Physics, University of California, Berkeley, CA 94720, USA}
\author{B.~Shank} \affiliation{Department of Physics, Stanford University, Stanford, CA 94305, USA}
\author{D.~Speller} \affiliation{Department of Physics, University of California, Berkeley, CA 94720, USA}
\author{K.M.~Sundqvist} \affiliation{Department of Physics, University of California, Berkeley, CA 94720, USA}
\author{A.N.~Villano} \affiliation{School of Physics \& Astronomy, University of Minnesota, Minneapolis, MN 55455, USA}
\author{B.~Welliver} \affiliation{Department of Physics, University of Florida, Gainesville, FL 32611, USA}
\author{D.H.~Wright} \affiliation{SLAC National Accelerator Laboratory/Kavli Institute for Particle Astrophysics and Cosmology, 2575 Sand Hill Road, Menlo Park 94025, CA}
\author{S.~Yellin} \affiliation{Department of Physics, Stanford University, Stanford, CA 94305, USA}
\author{J.J.~Yen} \affiliation{Department of Physics, Stanford University, Stanford, CA 94305, USA}
\author{J.~Yoo} \affiliation{Fermi National Accelerator Laboratory, Batavia, IL 60510, USA}
\author{B.A.~Young} \affiliation{Department of Physics, Santa Clara University, Santa Clara, CA 95053, USA}
\author{J.~Zhang} \affiliation{School of Physics \& Astronomy, University of Minnesota, Minneapolis, MN 55455, USA}

\collaboration{CDMS Collaboration}

\noaffiliation

\begin{abstract}

We report results of a search for Weakly Interacting Massive Particles (WIMPS) with the silicon detectors of the CDMS II experiment. This blind analysis of 140.2 kg-days of data taken between July 2007 and September 2008 revealed three WIMP-candidate events with a surface-event background estimate of $0.41_{-0.08}^{+0.20}(stat.)_{-0.24}^{+0.28} (syst.)$. Other known backgrounds from neutrons and \pb are limited to $< 0.13$ and $<0.08$ events at the 90\% confidence level, respectively. The exposure of this analysis is equivalent to 23.4 kg-days for a recoil energy range of 7--100 keV for a WIMP of mass 10 \gev. The probability that the known backgrounds would produce three or more events in the signal region is 5.4\%. A profile likelihood ratio test of the three events that includes the measured recoil energies gives a 0.19\% probability for the known-background-only hypothesis when tested against the alternative WIMP+background hypothesis. The highest likelihood occurs for a WIMP mass of 8.6~\gev and WIMP-nucleon cross section of 1.9$\times 10^{-41}$~cm$^2$.

\end{abstract}


\maketitle

%comment this out to remove line numbers
%\linenumbers

There is now overwhelming evidence that the bulk of the matter in our universe is in some nonluminous, nonbaryonic form \cite{Bertone:2004pz}.  
%Though there is broad consensus on the amount of this dark matter present in the cosmos, its composition has thus far eluded laboratory investigations. 
Weakly Interacting Massive Particles (WIMPs) \cite{Steigman:1984ac} %-- particles with masses between a few \gev and a few \tev and interaction strengths characteristic of the weak force -- 
form a leading class of candidates for this dark matter.  Particles of this type would be produced thermally in the early universe and are predicted by many theoretical extensions to the Standard Model of particle physics \cite{Lee:1977ua,Jungman:1995df,Bertone:2004pz}.  If WIMPs do constitute the dark matter in our galaxy, they may be detectable through their elastic scattering from nuclei in terrestrial particle detectors \cite{Goodman:1984dc}.  Numerous experimental groups have sought to detect such scattering events using a wide variety of technologies \cite{Bertone:2010}.

The Cryogenic Dark Matter Search (CDMS) collaboration identifies nuclear recoils (including those that would occur in WIMP interactions) using semiconductor detectors operated at 40~mK.  These detectors use simultaneous measurements of ionization and non-equilibrium phonons to identify such events among the far more numerous background of electron recoils. 

The low atomic mass of Si generally makes it a less sensitive target for spin-independent WIMP interactions relative to the larger coherent enhancement of the scattering cross section for heavy nuclei.  On the other hand, the lower atomic mass of Si is advantageous in searches for WIMPs of relatively low mass ($\sim$10~GeV/$c^{2}$) due to more favorable scattering kinematics.  New particles at such masses are generally disfavored in fits of models to precision electroweak data (\eg \cite{Baltz:2004aw,*Roszkowski:2007fd}), but viable models in this regime do exist (\eg \cite{Bottino:2003cz,Kaplan:2009ag,*Cohen:2009fz}).  Renewed interest in this mass range has been motivated by results from the DAMA/LIBRA \cite{Bernabei:2010yi}, CoGeNT \cite{Aalseth:2011}, and CRESST~\cite{CRESST:2012} experiments, which can be interpreted as evidence of low-mass WIMP scattering.

During 2003-2008 the collaboration operated CDMS II, an array of Ge and Si detectors located at the Soudan Underground Laboratory \cite{R118_PRD,Akerib:2004fq,Akerib:2005kh,Ahmed:2008eu,CDMSScience:2010,Ahmed:2009rh,Collaboration:2009ht}. In its final configuration, the CDMS II array consisted of 30 Z-sensitive ionization and phonon (ZIP) detectors: 19 Ge ($\sim$239 g each) and 11 Si ($\sim$106 g each), for a total of $\sim$4.6 kg of Ge and $\sim$1.2 kg of Si.  %Each CDMS~II detector is a semiconductor disk, 7.6 cm in diameter and 1 cm thick, instrumented to detect the phonons and ionization generated by particle interaction within the crystal \cite{R118_PRD}.  %One flat face of each detector is instrumented with four readout channels composed of superconducting transition-edge sensors (TESs) to detect nonequilibrium phonons.  The opposite flat face is divided into two concentric ionization electrodes: an inner (primary) electrode covering $\sim$85\% of the detector surface and an outer guard ring.  The latter defines a fiducial volume within each ZIP by identifying interactions near the detector rim, which may suffer from reduced ionization collection.  
We discriminate nuclear recoils from background electron recoils 
using the ratio of ionization to phonon recoil energy (ionization ``yield'').  Electron recoils that occur within $\sim$10~$\mu$m of a detector surface can exhibit reduced ionization collection.  These events are identified by phonon pulse-shape discrimination. Our overall misidentification rate of electron recoils is less than 1 in $10^6$.

We consider data from the Si detectors using the final four run periods of the full CDMS II detector installation, acquired between %October 2006 
July 2007 and September 2008. 
%The data were analyzed in two periods, one from October 2006 to July 2007 (Runs 123+124, henceforth c34) and another from July 2007 to September 2008 (Runs 125-128, henceforth c58). 
The Ge results from this data set have been described in previous publications \cite{CDMSScience:2010,Ahmed:low-thresh}.  Compared to Si data from the earlier CDMS II runs, described in \cite{CDMS-Si-c34:2013}, these data benefit from improved analysis and calibration techniques. Of the 11 Si detectors, three were excluded from the WIMP-search analysis: two due to wiring failures that led to incomplete collection of the ionization signal and one due to unstable response on one of its four phonon channels. %Two additional ``endcap'' detectors, located at the tops of two of the detector towers, were excluded from the c34 analysis due to the inability to tag multiple-scatters on the outer detector face. 
Periods of poor performance, as identified by a series of Kolmogorov-Smirnov tests, were also excluded from analysis. After all such exclusions, the data collected by the 8 Si detectors considered in this analysis represent a total exposure of 140.2 kg-days prior to the application of the WIMP candidate selection criteria.

The responses of these detectors to electron and nuclear recoils were calibrated using events from extensive exposures to \ba and \cf sources {\it in situ} at Soudan.  Electron recoils from the former were used to empirically characterize and correct for the dependence of phonon pulse shape on event position and energy.  %In the c34 analysis, events at large detector radii were excluded due to degraded performance of this correction technique; the correction technique was improved during the c58 analysis, and this cut was no longer employed.  
The 356~keV gamma ray from the \ba source has a $\sim4.2$~cm attenuation length in Si, and thus the Si detectors generally do not show a clear line at 356~keV. Their energy scales were calibrated using 356~keV events with total energies shared between the Si detector and a neighboring detector.  %The detectors' response to nuclear recoils was determined using neutron-scattering events from calibration with the \cf source.

\begin{figure}[tb]
\centering
\includegraphics[width=\columnwidth]{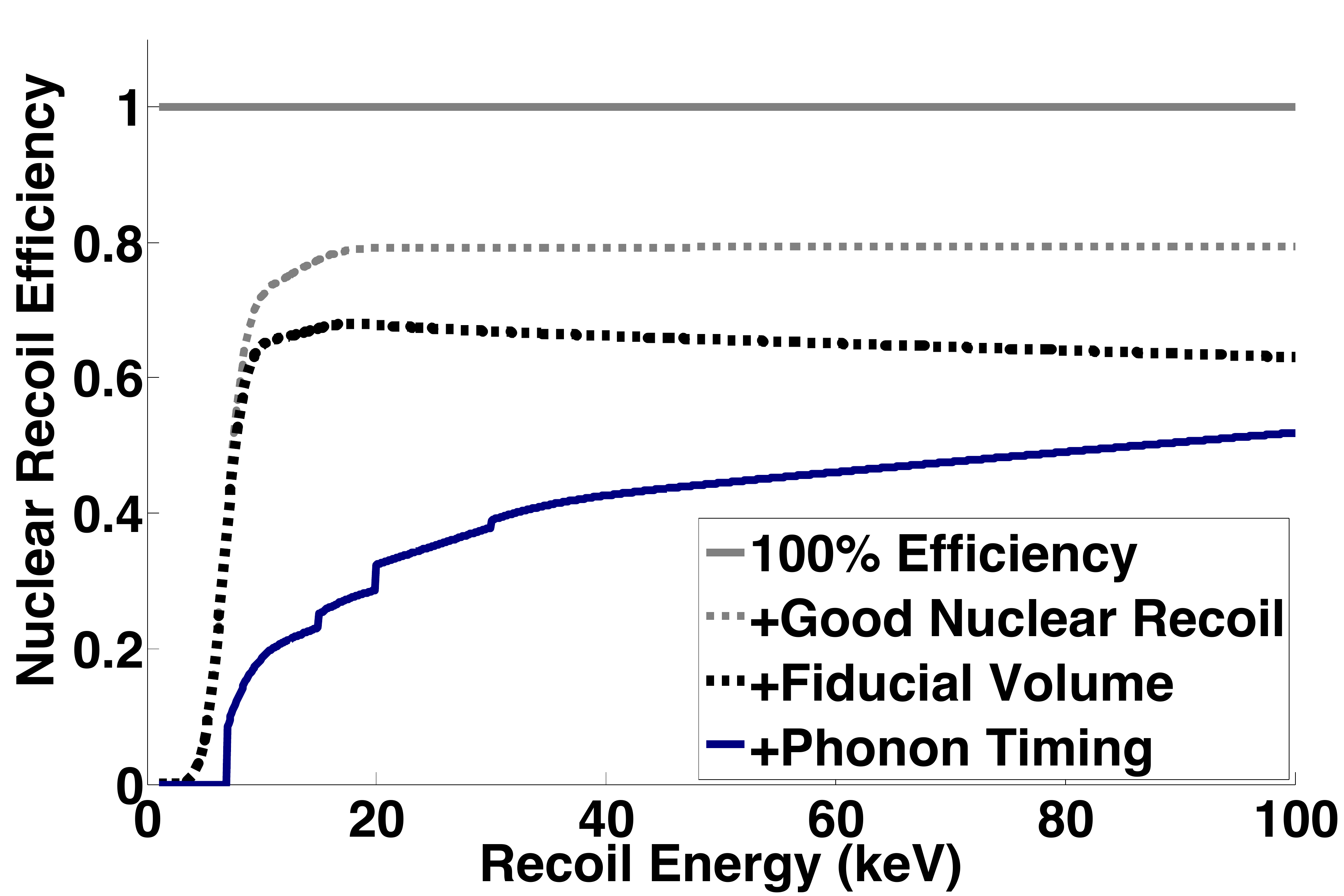}
\caption{Nuclear-recoil efficiency as function of recoil energy after application of each selection criterion shown. Each curve from top to bottom shows the cumulative effect of successive cuts on the data, with the second curve from the top (grey dashed) encompassing all data-quality cuts, trigger and ionization threshold efficiencies, and the nuclear-recoil yield band efficiency as measured on \cf neutrons.  The third curve (black dashed) adds the ionization radial cut to the above,  The bottom curve (solid blue) adds both the phonon timing criteria and the recoil-energy thresholds, and hence shows the overall efficiency of this analysis. The abrupt drops in acceptance at low recoil energies reflect the elevated energy thresholds chosen for some detectors. }
\label{fig:SiEff}
\end{figure}

WIMP-candidate events were identified by a series of selection criteria.  All WIMP selection criteria were defined using calibration data plus WIMP search data in which events in and near the WIMP candidate region were masked.
%Events in the WIMP-search data near the WIMP-candidate region (single events with yield within $\pm 3 \sigma$ of the nuclear recoil mean) were automatically removed from the data set during the analysis, and all WIMP-selection criteria were defined blindly using calibration and remaining WIMP-search data not removed by the blinding procedure. 
Thus, WIMP candidates had no impact on the definition of the selection criteria.  A WIMP candidate was required to have phonon and ionization signals above the noise in exactly one ZIP detector and to exhibit no coincident energy in the scintillating veto shield. Events in coincidence with the NuMI beam~\cite{Anderson:1998zza} were also vetoed. We demanded that any candidate event occur within the detector's fiducial volume, defined by requiring signal consistent with noise in the outer ionization electrode. The recoil energy of each candidate event had to lie below 100 keV and above a detector-dependent threshold ranging from 7 to 30 keV, chosen blindly using calibration data to keep the total expected leakage of bulk ERs into the NR band below 0.03 events. Candidate events were further required to lie $>4.5\sigma$ above the ionization channel noise as measured by randomly acquired triggers for each detector during each contiguous period of data taking ($\sim$24 hours ).

In yield, events were required to be within $+1.2\sigma$ and $-1.8\sigma$ from the mean of the nuclear recoil yield. Candidate events were also required to have phonon pulse timing consistent with a nuclear recoil. In order to take advantage of the fact that the timing parameters are better measured at high energies, the phonon timing data-selection cut was optimized in three energy bins: 7--20~keV, 20--30~keV, and 30--100~keV~\cite{McCarthy:2013}. Fig.~\ref{fig:SiEff} shows the nuclear-recoil efficiency i.e., the estimated fraction of nuclear recoils at a given energy that would be accepted by these signal criteria, measured using nuclear recoils from \cf calibration. The abrupt changes in efficiency are due to the different detector thresholds and changes to the timing cuts in the three energy bins. Signal acceptance was measured using nuclear recoils from \cf calibration.  After applying all selection criteria, the exposure of this analysis is equivalent to 23.4 kg-days over a recoil energy range of 7--100~keV for a WIMP of mass 10 \gev.% \emph{Change to 15 \gev exposure; I need to get c34 exposure at this mass to do so}.

\begin{figure}[tb]
\centering
\includegraphics[width=\columnwidth]{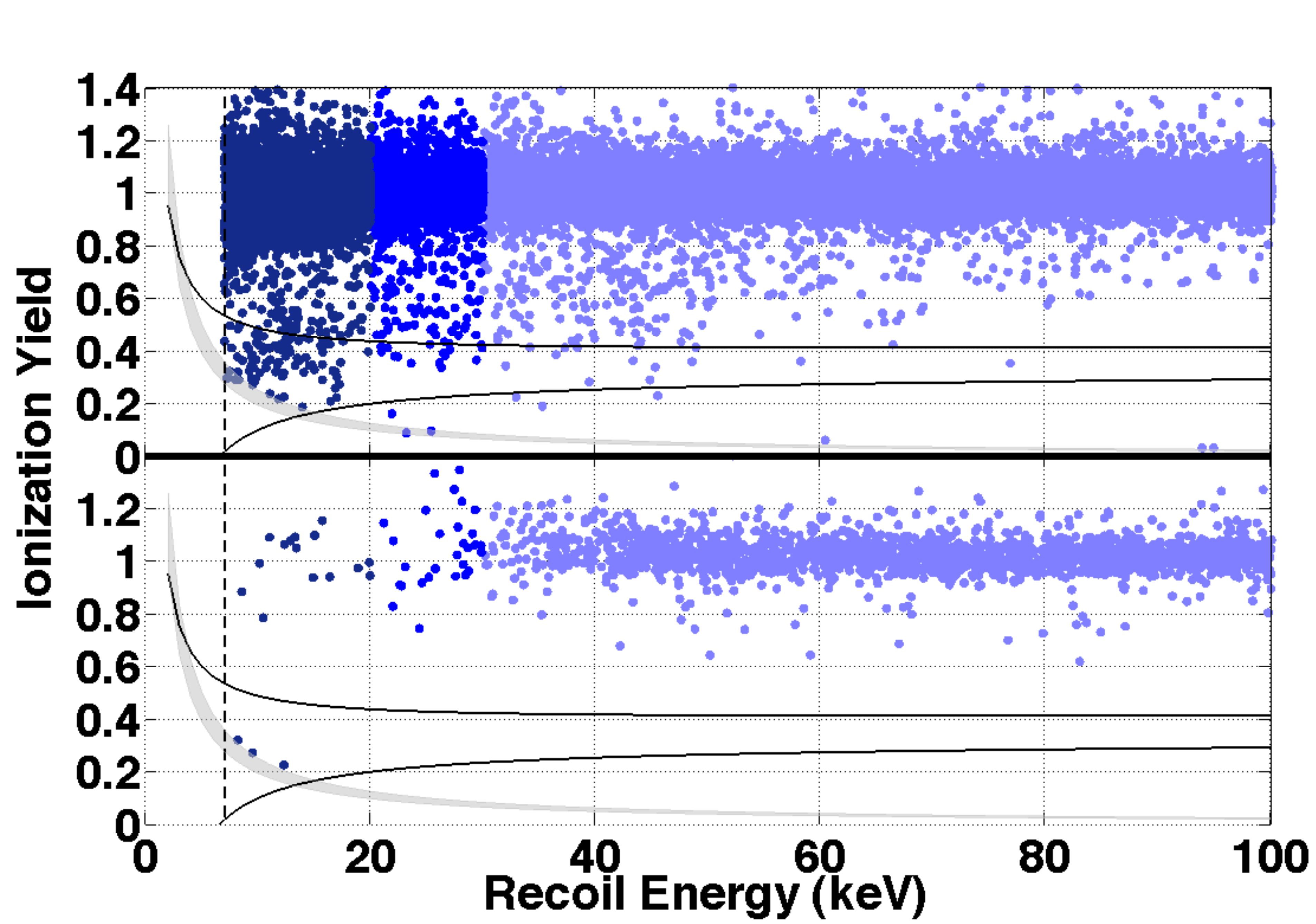}
\caption{Ionization yield versus recoil energy in all detectors included in this analysis for events passing all signal criteria except ({\it top}) and including ({\it bottom}) the phonon timing criterion.  The curved black lines indicate the signal region (-1.8$\sigma$ and +1.2$\sigma$ from the mean nuclear recoil yield) between 7 and 100~keV recoil energies for detector 3 in Tower 4, while the gray band shows the range of charge thresholds across detectors. Electron recoils in the detector bulk have yield near unity. The data are colored to indicate recoil energy ranges (dark to light) of 7--20, 20--30, and 30--100~keV to aid the interpretation of Fig.~\ref{fig:c34SiSignalBox}.}
\label{fig:c34SiTwoPanel}
\end{figure}

\begin{figure}[tb]
\centering
\includegraphics[width=\columnwidth]{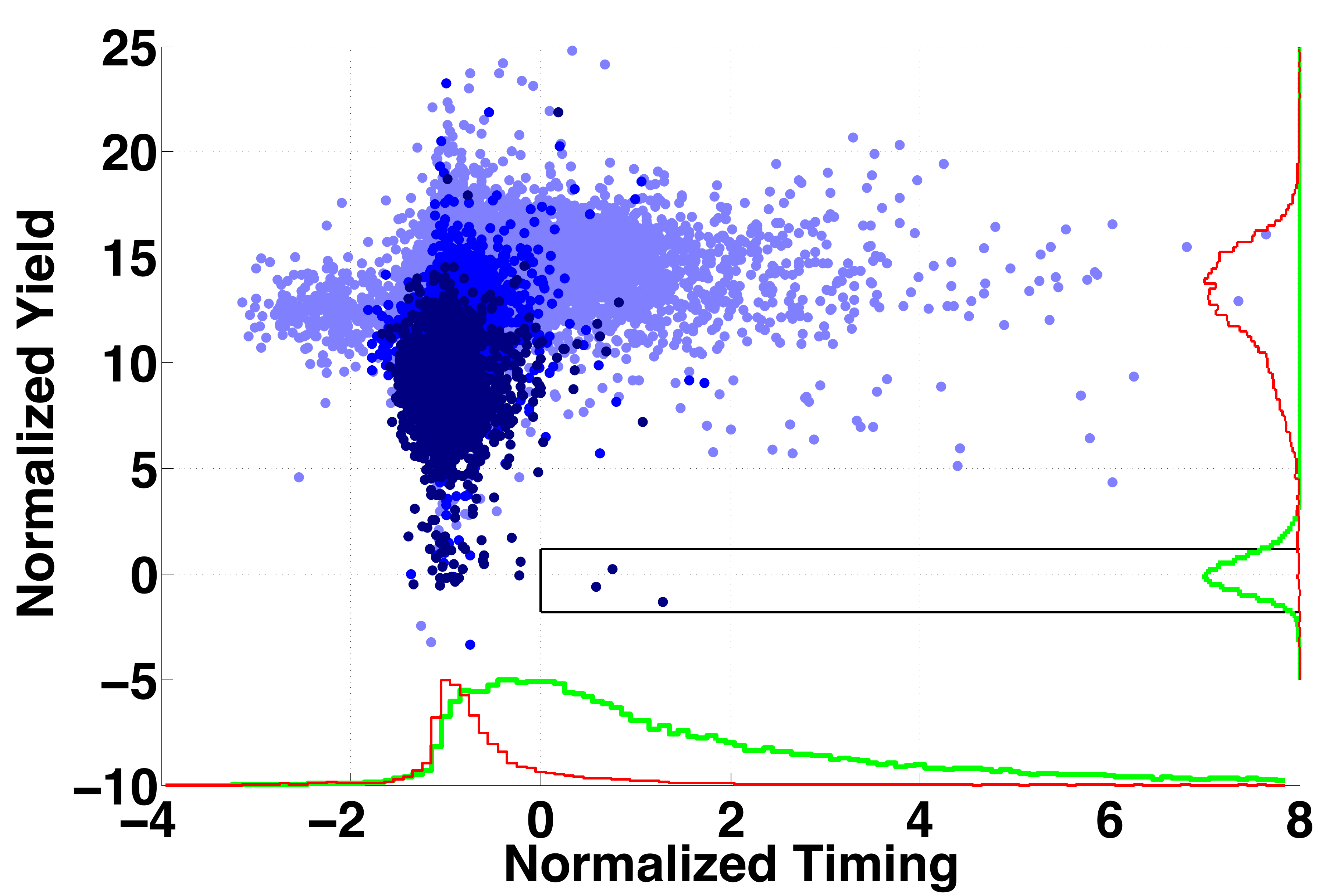}
\caption{Normalized ionization yield (standard deviations from the nuclear recoil band centroid) versus normalized phonon timing parameter (normalized such that the median of the surface event calibration sample is at -1 and the cut position is at 0) for events in all detectors from the WIMP-search data set passing all other selection criteria. The black box indicates the WIMP candidate selection region. The data are colored to indicate recoil energy ranges (dark to light) of 7--20, 20--30, and 30--100~keV. The thin red curves on the bottom and right axes are the histograms of the data, while the thicker green curves are the histograms of nuclear recoils from \cf calibration data; both are normalized to have the same arbitrary peak value.}
\label{fig:c34SiSignalBox}
\end{figure}

Neutrons from cosmogenic or radioactive processes can produce nuclear recoils that are indistinguishable from those from an incident WIMP.  Simulations of the rates and energy distributions of these processes using GEANT4~\cite{Geant4:2006} lead us to expect $<0.13$ false candidate events (90\% confidence level) in the Si detectors from neutrons for this exposure with all efficiencies included.

A greater source of background is the misidentification of surface electron recoils, which may suffer from reduced ionization yield and thus contribute events to the WIMP-candidate region; these events are termed ``leakage events''.  Prior to looking at the WIMP-candidate region (unblinding), the expected leakage was estimated using the rate of single scatter events with yields consistent with nuclear recoils from a previously unblinded Si dataset \cite{Filippini:2008} and the rejection performance of the timing cut measured on low-yield multiple-scatter events from \ba calibration data. 
%Prior to looking at the WIMP-candidate region (``unblinding''), the expected leakage was estimated as follows: a previously unblinded data set \cite{Filippini:2008} gave the rate of single scatter events in these detectors with yields consistent with nuclear recoils;  the fraction of such events that would pass the phonon timing cut was estimated from its performance on low-yield multiple-scatter events from \ba calibration data.
Two detectors used in this analysis were located at the end of detector stacks, so scatters on their outer faces could not be tagged as multiple scatters. The rate of surface events on the outer faces of these two detectors were estimated using their single-scatter rates from a previously unblinded dataset presented in \cite{Filippini:2008} and the multiples-singles ratio on the interior detectors. The final pre-unblinding estimate for misidentified surface electron-recoil event leakage into the signal band in the eight Si detectors was $0.47_{-0.17}^{+0.28}(stat.)$ events. This initial leakage estimate informed the decision to unblind. After unblinding, we developed a Bayesian estimate of the rate of misidentified surface events based upon the performance of the phonon timing cut measured using events near the WIMP-search signal region \cite{Filippini:2008,CDMS-Si-c34:2013}. Multiple-scatter events below the electron-recoil ionization-yield region from both \ba calibration and the WIMP-search data were used as inputs to this model. Because the WIMP-search sample is sparser compared to the calibration data, the combined estimates are more heavily weighted towards the calibration data leakage estimates.  Additionally the leakage estimate is corrected for the fact that the fraction of singles passing the timing cut is higher than the fraction of multiples by a factor of $1.7_{-0.6}^{+0.8}$, as measured on low-yield events outside of the nuclear recoil band. The systematic uncertainty on the leakage estimate comes from the uncertainty on this scale factor, the choice of prior in the Bayesian analysis, and the method used to reweigh the energy distribution of surface events from calibration data to reflect the distribution in WIMP search data. The final model predicts an updated surface-event leakage estimate of $0.41_{-0.08}^{+0.20}(stat.)_{-0.24}^{+0.28} (syst.)$ misidentified surface electron-recoil events in the eight Si detectors. Classical confidence intervals provided similar estimates \cite{schneeRuchlin}.

After all WIMP-selection criteria were defined, the signal regions of the Si detectors were unblinded on December 25, 2012. Three WIMP-candidate events were observed, with recoil energies of 8.2, 9.5, and 12.3~keV, on March~14, July~1, and September~6 of 2008, respectively. Two events were observed in Detector 3 of Tower 4, and the third was observed in Detector 3 of Tower 5. These detectors were near the middle of their respective tower stacks.
%Detailed examination of these events has revealed nothing suspicious about the events or their arrival times.
Fig.~\ref{fig:c34SiTwoPanel} illustrates the distribution of events in and near the signal region of the WIMP-search data set before ({\it top}) and after ({\it bottom}) application of the phonon timing criterion.  Fig.~\ref{fig:c34SiSignalBox} shows an alternate view of these events, expressed in ``normalized'' versions of yield and timing that are transformed so that the WIMP acceptance regions of all detectors coincide.

After unblinding, extensive checks of the three candidate events revealed no data quality or analysis issues that would invalidate them as WIMP candidates. The signal-to-noise on the ionization channel for the three events (ordered in increasing recoil energy) was measured to be 6.7$\sigma$, 4.9$\sigma$, and 5.1$\sigma$. A study on possible leakage into the signal band due to \pb recoils from $^{210}$Po decays found the expected leakage to be negligible with an upper limit of $<0.08$ events at the 90\% confidence level. The energy distribution of the \pb background was constructed using events in which a coincident $\alpha$ particle was detected in a detector adjacent to one of the 8 Si detectors used in this analysis. 

\begin{figure}
\centering
\includegraphics[width=\columnwidth]{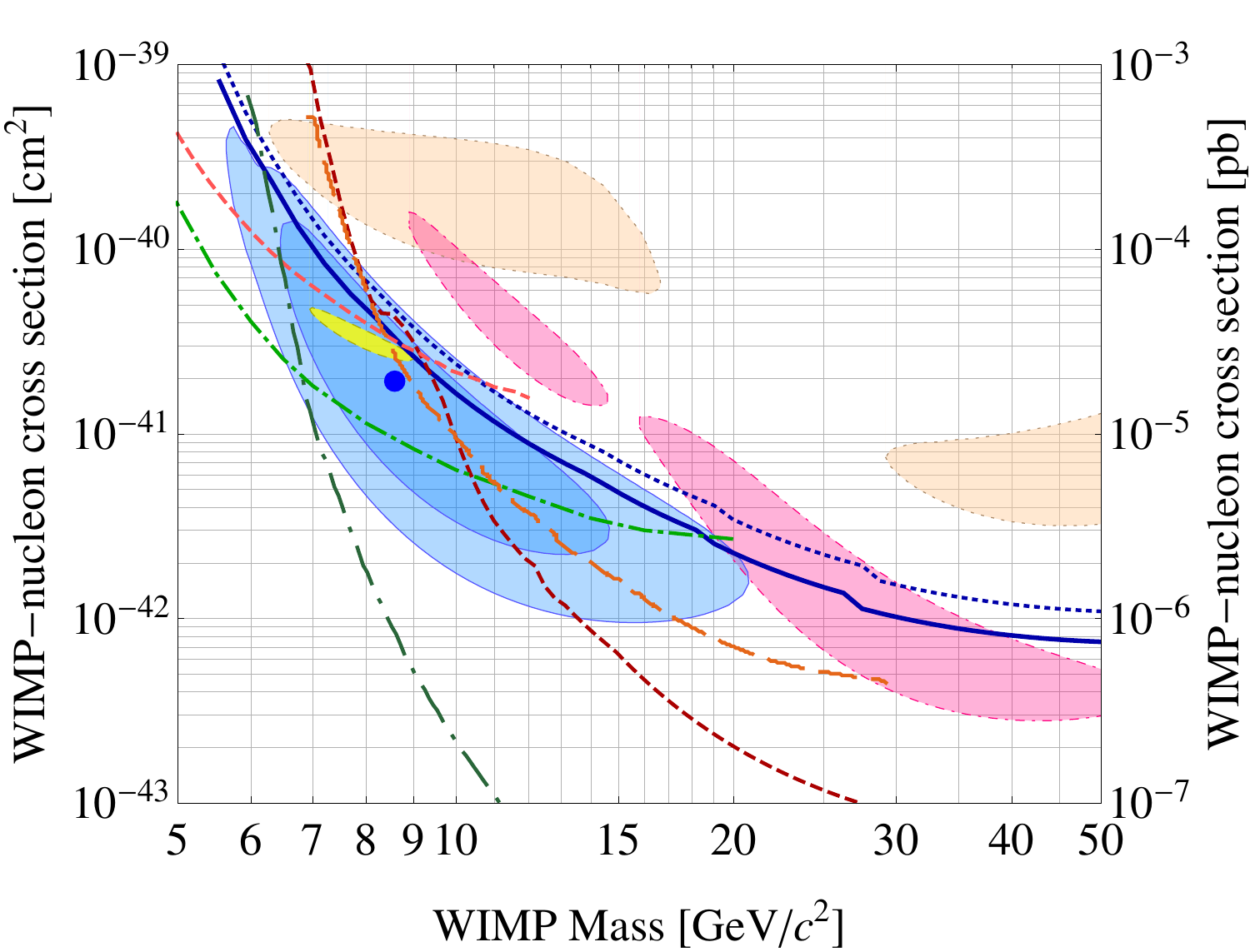}
\caption{Experimental upper limits (90\% confidence level) for the WIMP-nucleon spin-independent cross section as a function of WIMP mass. We show the limit obtained from the exposure analyzed in this work alone ({\it blue dotted line}), and combined with the CDMS II Si data set reported in \cite{Filippini:2008,CDMS-Si-c34:2013} ({\it blue solid line}).  Also shown are limits from the CDMS~II~Ge standard \cite{CDMSScience:2010} and low-threshold \cite{Ahmed:low-thresh} analysis  ({\it dark and light dashed red}), EDELWEISS low-threshold \cite{edel:2012} (\emph{long-dashed orange}), XENON10 S2-only \cite{XENON10S2} (\emph{dash-dotted green}), and XENON100 \cite{XENON100:2012} (\emph{long-dash-dotted green}). The filled regions identify possible signal regions associated with data from CoGeNT \cite{Aalseth:2012} ({\it dashed yellow}, 90\% C.L.), DAMA/LIBRA \cite{Bernabei:2010yi,Savage:2008er} ({\it dotted tan}, 99.7\% C.L.), and CRESST \cite{CRESST:2012,Brown:2012}  ({\it dash-dotted pink}, 95.45\% C.L.) experiments. 68\% and 90\% C.L. contours for a possible signal from these data alone are shown in light blue. The blue dot shows the maximum likelihood point at (8.6~GeV/c$^{2}$, $1.9\times10^{-41}$~cm$^{2}$).}
\label{fig:SILimits}
\end{figure}

This result constrains the available parameter space of WIMP dark matter models.  We compute upper limits on the WIMP-nucleon scattering cross section using Yellin's optimum interval method \cite{Yellin:2002xd}. We assume a WIMP mass density of $0.3$~\gev/cm$^3$, a most probable WIMP velocity with respect to the galaxy of $220$~\kms, a mean circular velocity of  Earth with respect to the galactic center of $232$~\kms, a galactic escape velocity of $544$~\kms~\cite{Smith:2006ym}, and the Helm form factor~\cite{Lewin:1995rx}. The effect of an annual modulation of the 10~\gev WIMP rate, found by integrating over the specific data-taking periods for this analysis with the above assumptions, introduces a $< 2\%$ shift downward in the cross sections of our results and is thus neglected. Fig.~\ref{fig:SILimits} shows the derived upper limits on the spin-independent WIMP-nucleon scattering cross section at the 90\% confidence level (C.L.) from this analysis and a selection of other recent results.  The present data set an upper limit of $2.4\times10^{-41}$ cm$^2$ for a WIMP of mass 10 \gev.  We are completing the calibration of the nuclear recoil energy scale using the Si-neutron elastic scattering resonant feature in the \cf exposures. This study indicates that our reconstructed energy may be 10\% lower than the true recoil energy, which would weaken the upper limit slightly.  Below 20~\gev the change is well approximated by shifting the limits parallel to the mass axis by $\sim7\%$, making the limits weaker at low masses. In addition, neutron calibration multiple scattering effects improve the response to WIMPs, thus shifting the upper limit down to a lower cross-section axis and making the limits stronger by $\sim 5\%$. 

%\begin{figure}
%\centering
%\includegraphics[width=\columnwidth]{figures/limitplot_allSi_Apr12.pdf}
%\caption{Experimental upper limits (90\% confidence level) for the WIMP-nucleon spin-independent cross section as a function of WIMP mass. We show the limit obtained from the exposure analyzed in this work alone ({\it black dots}), and the combined limit for the full CDMS II Si data set recorded at Soudan ({\it blue solid line}, this work plus \cite{Filippini:2008} and \cite{Ogburn:2008}).  Also shown are limits from the CDMS II Ge standard \cite{CDMSScience:2010} and low-threshold \cite{Ahmed:low-thresh} analysis  ({\it dark and light dashed red}), XENON10 S2-only \cite{XENON10S2} (\emph{light dash-dotted green}), and XENON100 \cite{Angle:2007uj} (\emph{dark dash-dotted green}). The filled regions identify possible signal regions associated with data from CoGeNT \cite{Kelso:2012} ({\it magenta}, 90\% C.L., as interpreted by Kelso \emph{et al.}~including the effect of a residual surface event contamination described in \cite{Aalseth:2011mod}), DAMA/LIBRA \cite{Bernabei:2010yi,Savage:2008er} ({\it yellow}, 99.7\% C.L.), and CRESST \cite{CRESST:2012}  ({\it brown}, 95.45\% C.L.) experiments. 68\% and 90\% C.L. contours for a possible signal from these data are shown in blue and cyan, respectively. The asterisk shows the maximum likelihood point at (8.6~GeV/c$^{2}$, $1.9\times10^{-41}$~cm$^{2}$).}
%\label{fig:SILimits}
%\end{figure}

A model of our known backgrounds, including both energy and expected rate distributions, was constructed for each detector and experimental run for each of the three backgrounds considered: surface electron recoils, neutron backgrounds, and \pb recoils.  Simulations of our background model yield a 5.4\% probability of a statistical fluctuation producing three or more events in our signal region.

This model of our known backgrounds was used to investigate the data in the context of a WIMP+background hypothesis. We performed a profile likelihood analysis, including the event energies, in which the background rates were treated as nuisance parameters and the WIMP mass and cross section were the parameters of interest. We profiled over probability distribution functions of the rate for each of our known backgrounds. The highest likelihood was found for a WIMP mass of 8.6~GeV/c$^2$ and a WIMP-nucleon cross section of 1.9$\times 10^{-41}$~cm$^2$.  The goodness-of-fit test of this WIMP+background hypothesis results in a p-value of 68\%, while the background-only hypothesis fits the data with a p-value of 4.5\%. A profile likelihood ratio test finds that the data favor the WIMP+background hypothesis over our background-only hypothesis with a p-value of 0.19\%. 
Though this result favors a WIMP interpretation over the known-background-only hypothesis, we do not believe this result rises to the level of a discovery.

Fig.~\ref{fig:SILimits} shows the resulting best-fit region from this analysis (68\% and 90\% confidence level contours) on the WIMP-nucleon cross-section vs.~WIMP mass plane. The 90\% C.L. exclusion regions from CDMS II's Ge and Si analyses and EDELWEISS low-threshold analysis cover part of this best-fit region, but the results are overall
statistically compatible. While there is some tension with the upper limits from the XENON10 experiment, the XENON100 experiment significantly constrains this parameter space under standard assumptions about the WIMP velocity distribution and WIMP-nucleus interactions. Additional, planned studies of these CDMS II Si data with reduced threshold may provide additional insight into a WIMP interpretation of these data. Future experiments with Si-based detectors that would be sensitive to WIMPs in this region of parameter space are also under consideration by the SuperCDMS collaboration.

The CDMS collaboration gratefully acknowledges the contributions of numerous 
engineers and technicians; we would like to especially thank Dennis Seitz, 
Jim Beaty, Bruce Hines, Larry Novak, 
Richard Schmitt and Astrid Tomada.  In addition, we gratefully acknowledge assistance 
from the staff of the Soudan Underground Laboratory and the Minnesota Department of Natural Resources. 
This work is supported in part by the 
National Science Foundation (Grant Nos.\ AST-9978911, NSF-0847342, NSF-1151869, PHY-0542066, 
PHY-0503729, PHY-0503629,  PHY-0503641, PHY-0504224, PHY-0705052, PHY-0801708, PHY-0801712, PHY-0802575, PHY-0847342, PHY-0855299, PHY-0855525, PHY-1102795, and PHY-1205898), by
the Department of Energy (Contracts DE-AC03-76SF00098, DE-FG02-92ER40701, DE-FG02-94ER40823, DE-FG03-90ER40569, and DE-FG03-91ER40618, and DE-SC0004022), by the Swiss National 
Foundation (SNF Grant No. 20-118119), by NSERC Canada (Grants SAPIN 341314 and SAPPJ 386399), and by MULTIDARK CSD2009-00064 and FPA2012-34694. Fermilab is operated by Fermi Research Alliance, LLC under Contract No. De-AC02-07CH11359, while SLAC is operated under Contract No.  DE-AC02-76SF00515 with the United States Department of Energy.

\bibliography{R125-8_Si}
\bibliographystyle{apsrev4-1}

\end{document}